\documentclass[aps,twocolumn,superscriptaddress,footinbib,longbibliography,prb]{revtex4-1} 

\usepackage{colortbl}
\usepackage{color}
\usepackage{amsfonts,amsmath,amssymb}
\usepackage[dvipsnames]{xcolor}
\usepackage{tabularx,hhline,tikz,multirow,array}
\usepackage{bm,enumitem}
\usepackage{dcolumn,stmaryrd}
\usepackage[bookmarksnumbered,pdfpagelabels=true,plainpages=false,colorlinks=true,linkcolor=blue,citecolor=blue,urlcolor=blue]{hyperref}
\usepackage{comment}
\definecolor{red}{rgb}{1,0,0}
  
\newcommand{\bs}[1]{\boldsymbol{#1}}

\begin{document}
\title{Magnons in the strained Heisenberg-Kitaev magnet}
\author{Miguel Letelier}
\affiliation{Departamento de Física, Facultad de Ciencias Físicas y Matemáticas, Universidad de Chile, Casilla 487-3 Santiago, Chile}
\author{Roberto E. Troncoso}
\affiliation{Departamento de Física, Facultad de Ciencias, Universidad de Tarapacá, Casilla 7-D, Arica, Chile}
\author{Nicolas Vidal-Silva}
\email{nicolas.vidal@ufrontera.cl}
\affiliation{Departamento de Ciencias F\'isicas, Universidad de La Frontera, Casilla 54-D, Temuco, Chile}

\date{\today}

\begin{abstract}
The properties of magnons hosted in strained Heisenberg-Kitaev magnets are investigated using numerical and analytical calculations. Considering that deformation fields modulate the coupling parameters, we find a general expression for the weakly strained magnon Hamiltonian that depends on the (symmetric) strain tensor. We numerically tested our results in finite nanoribbon structures. We found that uniaxial deformations make the bulk bands more dispersive while topologically protected in-gap edge modes become nonreciprocal at the boundary of the Brillouin Zone. On the other hand, when applying a twist deformation, the simultaneous modulation of both Heisenberg and Kitaev parameters enables the apparition of flat bands, promoting the presence of non-propagative topologically protected magnonic edge states, whose properties strongly depends on the strain strength. In addition, the characteristic localization of magnon edge modes is preserved, which allows for testing the robustness of the bulk-boundary correspondence under lattice deformations. Our results contribute to a major understanding of Heisenberg-Kitaev magnets and how applying different strains allows for precise control over magnon properties.
\end{abstract}

\date{\today}

\maketitle

\section{Introduction}
Straintronics, the uses of mechanical strain to control and engineering emergent  physical effects in condensed matter systems, have gained significant attention during the past decade 
\cite{bukharaev2018straintronics,miao2021straintronics,roy2011hybrid,bandyopadhyay2024perspective}. Under strain the geometry of the crystalline lattice is deformed, inducing changes in the microscopic properties of material systems. This mechanism has been widely employed, e.g., in semiconductors \cite{sun2007physics,sun2009strain}, ferroelectric materials \cite{schlom2007strain,haeni2004room}, topological insulators \cite{liu2014tuning,aramberri2017strain,agapito2013novel}, superconductors \cite{ruf2021strain,pogrebnyakov2004enhancement,ekin1984strain}, optoelectronics \cite{du2021strain,li2020recent,khan2020tunable} and magnetic materials \cite{webster2018strain,cenker2022reversible,lei2013strain}. 
A remarkable example is graphene \cite{naumis2017electronic,ribeiro2009strained,si2016strain}, where the application of specific strain give rise to elastic gauge fields, leading to the formation of pseudo Landau levels \cite{suzuura2002phonons,guinea2010energy,guinea2010generating,vozmediano2010gauge,levy2010strain,guinea2010energy,guinea2010generating}, the shifts of Dirac cones \cite{choi2010effects,ribeiro2009strained,cocco2010gap}, changes in the fermi velocity of Dirac electrons \cite{choi2010effects}, and enhancements of the electron-phonon coupling \cite{si2013first}. 

Magnetic materials have set the ground for several spin-based potential applications.  Specifically, magnons, their low-energy excitations, have been proposed as efficient spin carriers as they lack Joule losses when propagating, converting them as a promising key building block for future low-energy applications \cite{flebus20242024,chumak2015magnon,yuan2022quantum,chumak2014magnon}. In this regard, the precise control of magnon properties is essential to achieve such a goal. A natural step towards the desired control of magnons relies on the possibility of employing strain engineering on distinct magnetic materials. Magnons in strained magnets have been focused changes of magnon energy \cite{brehm2024magnon,kim2022giant}, the magnon relaxation time \cite{wang2024strain}, strain-mediated magnon Landau Levels (LLs)\cite{nayga2019magnon,sun2021magnon,wei2024strain,sun2021quantum,ferreiros2018elastic}, topological phase transitions \cite{owerre2018strain,vidal2022time,soenen2023tunable}, and strain-controlled magnon spin currents \cite{zhou2022piezoelectric,kondo2022nonlinear,van2024magnon}. The physical origin of such effects corresponds to strain-induced modifications over the exchange interaction due to the dependency over separation between adjacent spins.

Special attention should be paid to frustrated magnets, where the competing spin interactions are bond-dependent, as in the case of the Kitaev model. Although there is no evidence for materials being stable solely containing Kitaev interaction, it has been pointed out as responsible for several magnetic properties of distinct materials while competing with Heisenberg-like exchange interactions, coined thus the concept {Heisenberg-Kitaev} (H-K) magnets. Some materials relying on this kind of materials are Na$_2$IrO$_3$, $\alpha$-RuCl$_3$, or CrI$_3$\cite{aguilera2020topological,lee2020fundamental} which have been shown to present a variety of magnetic ordering \cite{smit2020magnon,li2017kitaev,maksimov2022easy,li2022thermal} according to the values for the magnetic parameters. When perturbed, such states can host magnon excitations \cite{li2017kitaev,ozel2019magnetic}, whose nonlinear aspects have been also addressed \cite{smit2020magnon,maksimov2022easy} to show quantum and damping effects. Magnons in H-K magnets have shown to possess a topological origin \cite{aguilera2020topological,li2022thermal,li2023topological,joshi2018topological,lee2020fundamental,zhang2021interplay}, which indeed form chiral magnon-polaron excitations when interacting with phonons \cite{mella2024chiral}.

The exploration of strain-induced effects on Heisenberg-Kitaev materials has been addressed mainly in the context of Majorana fermions \cite{fremling2022sachdev,agarwala2021gapless} and the desired control of both Heisenberg ($J$) and Kitaev ($K$) parameters \cite{yadav2018strain,kim2013strain}, leading to the possibility of reaching the quantum spin-liquid state for diminishing $J$, which is exactly solvable \cite{takagi2019concept}. Much less explored are the magnon properties under the action of strains in the mentioned materials, whose effect remains unclear. In this regard, a recent work by Soenen et al. [\citenum{soenen2023tunable}] has numerically explored by DFT methods the magnon topology in a CrI$_3$ monolayer described by including both Kitaev and DM terms under the action of distinct strains, giving rise to a topological phase transition. However, the sole presence of the H-K system (DM null) was not systematically studied. Thus, although the magnon properties have been explored in strained two-dimensional magnets, they mainly focused on the strain-induced exchange modulation. However, in H-K magnets, the bond-dependent Kitaev interaction has not been considered as deformed when strained. Since the role that both interactions play on magnon properties are essentially different, their modulation by strains might have significant consequences on magnon properties. More specifically, the effect of lattice deformations on the intrinsic topological properties of magnons hosted in H-K magnets is addressed here.\\

In this work, we explore the properties of magnons hosted in H-K magnets when subjected to distinct external strains, putting special attention on the presence of protected edge states and their interplay with lattice deformations. Since the interactions originated from Kitaev and exchange terms depend on the relative spin positions, we consider that both interactions are affected by the application of such deformations. 

The paper is laid out as follows. In section II, we present the spin Hamiltonian and the corresponding spin-elastic model. In section III, we formulate the theory of spin fluctuations in the strained H-K model. Section IV is devoted to the application of our model to semi-infinite nanoribbons by exerting both uniaxial and twisting deformations. 
Finally, in section V, we provide the conclusions and discuss the implications raised in our work.

\begin{figure}
    \centering
    \includegraphics[width = 8.5cm]{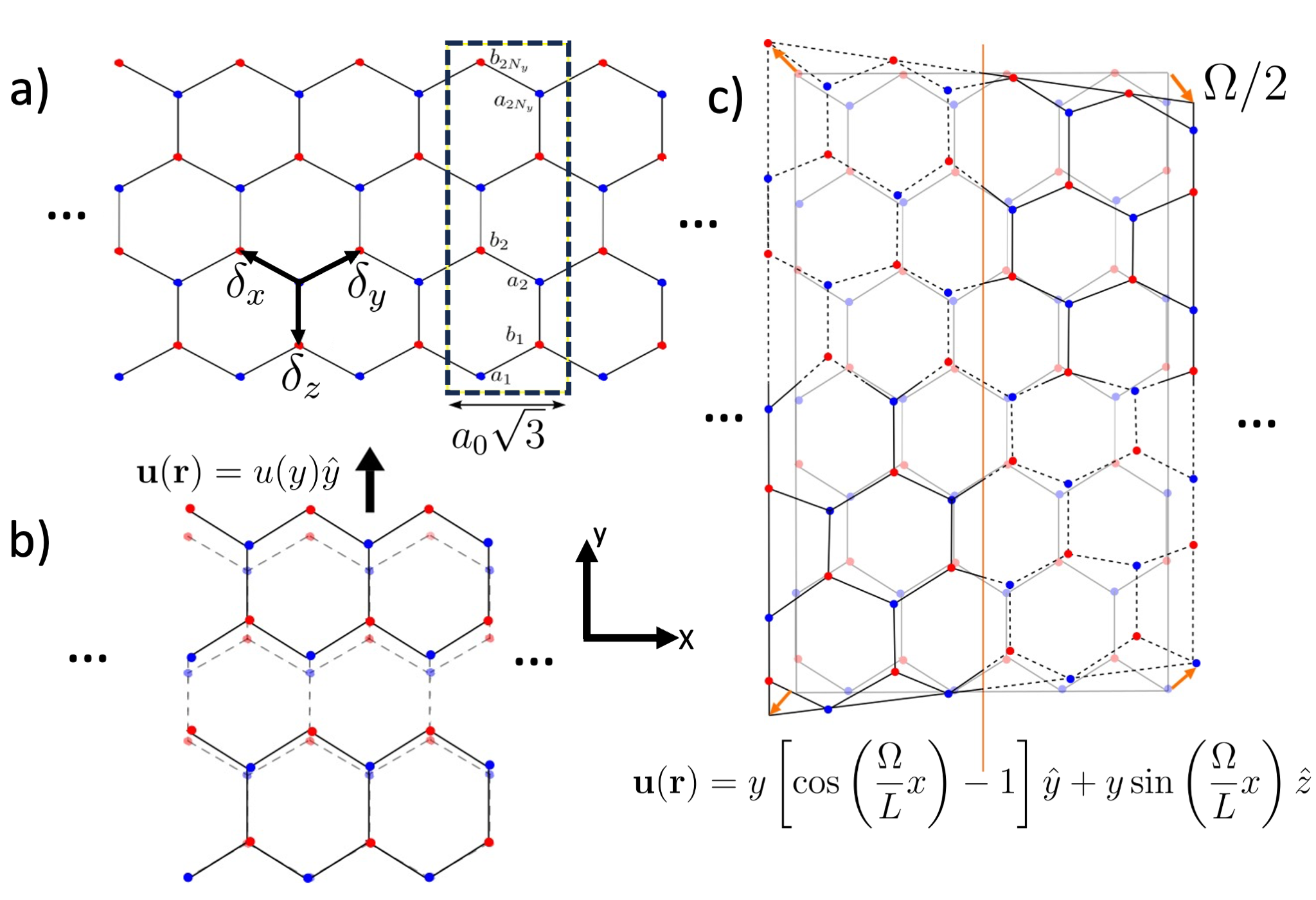}
    \caption{Schematic setups for the strained hexagonal lattices. In panel a), we show the unstrained nanoribbon, periodic and finite along the $x-$ and $y-$directions, respectively. The unit cell and the corresponding link vectors $\boldsymbol{\delta}_{x}=a_0\left(-\sqrt{3}/2,1/2\right), \boldsymbol{\delta}_{y}=a_0\left(0,-1\right)$, and $\boldsymbol{\delta}_{z}=a_0\left(\sqrt{3}/2,1/2\right)$, are highlighted. At b) the strained nanoribbon subjected to the uniaxial deformation field ${\bs u}(\bs r)=c_{\chi}y    A^2\hat{y}$ is shown, while at c) the crystal structure twisted (respect the $y-$axis) by the field ${\bs u}(\bs r)=y\left(\cos\left(\Omega x/L\right)-1\right)\hat{y}+y\sin\left(\Omega x/L\right)\hat{z}$.}
    \label{fig:fig1}
\end{figure}

\section{Spin-elastic model}
We consider a two-dimensional ferromagnetic system of localized spins lying on a honeycomb lattice, described by the Heisenberg-Kitaev (H-K) model whose spin Hamiltonian is given by,
\begin{align}\label{eq:GeneralSpinHamiltonian}
\mathcal{H}[{\bs S}] = -\sum_{\langle ij\rangle}J_{ij}&\nonumber{\boldsymbol{S}}_i\cdot{\boldsymbol{S}}_j-B\sum_{i}S^z_i\\
+&2\sum_{\langle ij\rangle_{\gamma}}K^{\gamma}{S}^{\gamma}_i{S}^{\gamma}_j+2\sum_{\langle ij\rangle_{\gamma}}\Gamma^{\gamma}{S}^{\alpha}_i{S}^{\beta}_j,
\end{align}
where $J_{ij}$ corresponds to the nearest neighbors ferromagnetic exchange interaction and $B$ is the applied magnetic field along the $z$-direction. The last two terms comprise the bond-dependent Kitaev and anisotropic exchange interaction \cite{joshi2018topological,mcclarty2018topological}, respectively, with $\gamma$ labeling each bond and the indices $\{\alpha,\beta\}$ running over spin components, as depicted in Fig. \ref{fig:fig1}. The Hamiltonian \ref{eq:GeneralSpinHamiltonian} harbor a complex phase diagram of magnetic ground states, e.g., zigzag antiferromagnetic order, stripy, ferromagnetic, among others  \cite{Janssen2019,Janssen2020}. In this work, we focus on a out-of-plane ferromagnetic order. Considering $K^{\gamma}=K$ and the parametrization $K=\sin\phi$ and $J=\cos\phi$, the range parameters that ensures its stability lies within the region $0.85\pi<\phi<3\pi/2$ for $\Gamma=0$ (and moderate values) \cite{joshi2018topological,mcclarty2018topological,janssen2016honeycomb}. 

Under strain of the crystal the relative change in the length of bonds connecting different points give rise to the interaction between spin and elastic degrees of freedom. The coupling generates through distortion of the crystal, which is described by a displacement field $\boldsymbol{u}({\bs R}_i)$ of the lattice site at position ${\bs R}_i$, that modifies the equilibrium position, $\boldsymbol{R}_i$, of atoms as $\boldsymbol{r}_i=\boldsymbol{R}_i+\boldsymbol{u}({\bs R}_i)$. The magnetoelastic coupling, comprising the exchange and bond-dependent interactions and represented by the symbol $\mathcal{X}$, depend on the relative position of spins. Therefore, the coupling is approximated by $\mathcal{X}_{ij}\equiv\mathcal{X}(\boldsymbol{r}_i-\boldsymbol{r}_j)|_{j=i+{\bs\delta}_{\eta}}\approx \mathcal{X}+c_{\mathcal{X}}\boldsymbol{\delta}_{\eta}\cdot ({\bs u}({\bs R}_i)-{\bs u}({\bs R}_{i+\boldsymbol{\delta}_{\eta}}))$ under small distortions of the lattice, where $\mathcal{X}\equiv\mathcal{X}(\boldsymbol{R}_i-\boldsymbol{R}_j)$ and $\eta$ running along the nearest neighbors. For large lattice, the displacement is considered as a smooth function of the coordinates. Spatial changes of the magnetic couplings are therefore approximated by
\begin{align}\label{eq:magcoupling}
\mathcal{X}_{ij}\approx \mathcal{X}-c_{{\cal X}}\,h_{\eta}(\bm{r}),
\end{align}
where $c_{{\cal X}} \boldsymbol{\delta}_{\eta} \approx \nabla_{\eta} {\cal X}$ and the field 
$h_{\eta}(\boldsymbol r)=\delta^a_{\eta}\delta^b_{\eta}u_{ab}(\bs r)$, where $u_{ab}=\left(\partial_au_b+\partial_bu_a\right)/2$ is the strain tensor and $\{a,b\}$ denoting the cartesian coordinates. Note that the field $h_{\eta}(\boldsymbol r)$ is solely sensitive to the properties of geometrical distortions such as, direction, magnitude, and spatial dependence encoded in the displacement field. At Fig. \ref{fig:fig1} we display some examples; the uniaxial and twisted strain. On the other hand, the factor $c_{{\cal X}}={\cal X}'$ is a measure of the strength of magnetoelastic coupling, that derives from the microscopic details of the actual magnetic coupling. In summary, the total spin-lattice Hamiltonian now reads as ${\cal H}={\cal H}_0+{\cal H}_d$, being ${\cal H}_0$ the unstrained Hamiltonian and ${\cal H}_d$ the magnetoelastic contribution, which is determined by the model for the crystal deformations under consideration. As it is expected, the magnetoelastic term at the total Hamiltonian modifies the energy landscape for magnetic phases, inducing transitions through various magnetic ordering. However, we will concentrates on the low-energy spin fluctuations of the magnetic phase for moderate magnetoelastic energy.

\section{Magnonic fluctuations}
We now introduce the quantum spin fluctuations around the out-of-plane ferromagnetic order. Within the linear spin wave theory, we expand the spin operators about the collinear state using the Holstein-Primakoff (HP) transformation \cite{holstein1940field} $S^{+}_{\boldsymbol{r}}\approx \sqrt{2S}a_{\boldsymbol{r}}$, $S^{-}_{\boldsymbol{r}} \approx \sqrt{2S}a_{\boldsymbol{r}}^{\dagger}$, and $S^z_{\boldsymbol{r}} = S-a^{\dagger}_{\boldsymbol{r}}a_{\boldsymbol{r}}$, where $a^{\dagger}_{\boldsymbol{r}}$ ($a_{\boldsymbol{r}}$) represents an operator that creates (annihilates) a magnon excitation at position $\boldsymbol{r}$. Next, we replace the HP transformation in the total Hamiltonian, defining the operators $a_{\bs r}$ and $b_{\bs r}$ on the sublattice $\mathcal{A}$ and $\mathcal{B}$, respectively, and keep only the bilinear terms to obtain a tight-binding magnon Hamiltonian ${H}_m$. After Fourier transform the bulk magnon Hamiltonian in momentum space reads 
\begin{align}\label{eq:magnonH}
{H}_m=\frac{S}{2}\sum_{\boldsymbol k\boldsymbol k'}\Psi^{\dagger}_{\boldsymbol k}{\cal M}_{\boldsymbol{kk'}}\Psi_{\boldsymbol k'},
\end{align}
with the field operator $\Psi_{\boldsymbol{k}}=(a_{\boldsymbol{k}},b_{\boldsymbol{k}},a^{\dagger}_{-\boldsymbol{k}},b^{\dagger}_{-\boldsymbol{k}})^T$. The Hamiltonian \ref{eq:magnonH} represents the low-energy spin fluctuations of weakly strained H-K ferromagnets, which is non-local in reciprocal space and favors magnons processes with different momentum. The $4\times 4$-matrix ${\cal M}$ is given by 
\begin{align}
\label{eq:defHamiltonian}
{\cal M}_{\boldsymbol k\boldsymbol k'}
&=\left(
\begin{array}{cc}
{\cal A}_{\boldsymbol k\boldsymbol k'}   & {\cal B}_{\boldsymbol k\boldsymbol k'}\\
{\cal{B}}^{\dagger}_{\boldsymbol k'\boldsymbol k}   & {\cal A}_{\boldsymbol k\boldsymbol k'} 
\end{array}
\right),
\end{align}
where the components of the $2\times 2$-matrix ${\cal A}$ and ${\cal B}$ read 
\begin{align*}
{\cal A}_{\boldsymbol k\boldsymbol k'}^{11}=&\kappa_0 \delta_{\boldsymbol k\boldsymbol k'}+{\cal K}^{(3)}_{\boldsymbol k\boldsymbol k'}\\
{\cal A}_{\boldsymbol k\boldsymbol k'}^{22}=&\kappa_0 \delta_{\boldsymbol k\boldsymbol k'}+{\cal K}^{(3)}_{\boldsymbol k\boldsymbol k'} e^{i(\boldsymbol k-\boldsymbol k')\cdot{\boldsymbol \delta}_z}\\
{\cal A}_{\boldsymbol k\boldsymbol k'}^{21}=&\kappa_{1,\boldsymbol{k}}  \delta_{\boldsymbol k\boldsymbol k'}+{\bar {\cal J}}_{\boldsymbol k'\boldsymbol k}+{\cal K}^{(2)}_{\boldsymbol k'\boldsymbol k}\\
{\cal B}_{\boldsymbol k\boldsymbol k'}^{21}=&\kappa_{2,\boldsymbol{k}}\delta_{\boldsymbol k\boldsymbol k'} +{\bar{\cal K}}^{(1)}_{-\boldsymbol k',\boldsymbol k}+{\cal G}_{-\boldsymbol k',\boldsymbol k}\\
{\cal B}_{\boldsymbol k\boldsymbol k'}^{12}=&\kappa_{2,-\boldsymbol{k}}\delta_{\boldsymbol k\boldsymbol k'} +{\bar{\cal K}}^{(1)}_{\boldsymbol k,-\boldsymbol k'}+{\cal G}_{\boldsymbol k,-\boldsymbol k'}
\end{align*}
and ${\cal A}_{\boldsymbol k\boldsymbol k'}^{12}=\bar{\cal A}_{\boldsymbol k\boldsymbol k'}^{21}$, ${\cal B}_{\boldsymbol k\boldsymbol k'}^{11}={\cal B}_{\boldsymbol k\boldsymbol k'}^{22}=0$. For the unstrained lattice the magnon Hamiltonian is determined by the coefficients, $\kappa_0=-(3J+2K)$, $\kappa_{1,\boldsymbol{k}}=\sum_\eta (J+K)e^{-i\boldsymbol{k}\cdot\boldsymbol{\delta}_\eta}
-Ke^{-i\boldsymbol{k}\cdot\boldsymbol{\delta}_z}$, and $\kappa_{2,\boldsymbol{k}}=i\Gamma e^{-i\boldsymbol{k}\cdot\boldsymbol{\delta}_z}+K\left(e^{-i\boldsymbol{k}\cdot\boldsymbol{\delta}_x}-e^{-i\boldsymbol{k}\cdot\boldsymbol{\delta}_y}\right)$.
Moreover, the contributions to the model from the strained lattice appear through the fields, 
\begin{align*}
{\cal J}_{\boldsymbol k\boldsymbol k'}&=\frac{c_J}{N}\sum_{\boldsymbol r,\eta} h_{\eta}(\boldsymbol r)e^{-i({\boldsymbol k}-{\boldsymbol k'})\cdot{\boldsymbol r}}e^{i{\boldsymbol k'}\cdot{\boldsymbol \delta}_{\eta}},\\
{\cal K}^{(1)}_{\boldsymbol k\boldsymbol k'}&=\frac{c_{K^x}}{N}\sum_{\bs r}e^{i\boldsymbol r\cdot(\boldsymbol k + \boldsymbol k')}\left(e^{i\boldsymbol k'\cdot\boldsymbol\delta_x}h_x(\boldsymbol r)-e^{i\boldsymbol k'\cdot\boldsymbol\delta_y}h_y(\boldsymbol r)\right),\\
{\cal K}^{(2)}_{\boldsymbol k\boldsymbol k'}&=\frac{c_{K^y}}{N}\sum_{\bs r}e^{-i\boldsymbol r\cdot(\boldsymbol k-\boldsymbol k')}\left(e^{i\boldsymbol k'\cdot\boldsymbol\delta_x}h_x(\boldsymbol r)+e^{i\boldsymbol k'\cdot\boldsymbol\delta_y}h_y(\boldsymbol r)\right),\\
{\cal K}^{(3)}_{\boldsymbol k\boldsymbol k'}&=2\frac{c_{K^z}}{N}\sum_{\bs r}e^{-i\boldsymbol r\cdot(\boldsymbol k-\boldsymbol k')}h_z(\boldsymbol r),\\
{\cal G}_{\boldsymbol k\boldsymbol k'}&=-i\frac{c_{\Gamma}}{N}\sum_{\bs r}e^{-i\boldsymbol{r}\cdot(\boldsymbol k +\boldsymbol k')}h_z(\boldsymbol r)e^{-i\boldsymbol k'\cdot\boldsymbol\delta_z},
\end{align*}
where $c_{\cal X}$ stands for the different sources of magnetoelastic coupling. Note that near the Dirac points the effects of elastic deformations are captured by the field ${\bs h}(\bs r)$, which resembles the emergence of elastic gauge fields \cite{nayga2019magnon,sun2021magnon,wei2024strain,sun2021quantum,ferreiros2018elastic,vozmediano2010gauge}. To solve the eigenvalue problem stated by the Hamiltonian \ref{eq:magnonH}, and the functional form of the deformation fields, approximations must be employed. However, the diagonalization can be exact for deformations linear in positions. Since our focus is on the effects of strains in the magnon spectrum, such as pseudo Landau levels, we will consider semi-infinite nanoribbons under uniaxial and twist deformations. In addition, the calculation of magnon spectrum will be complemented by the spectral function,
\begin{align}
A(\omega_,k_x)=-\frac{1}{\pi}\sum_{y}\lim_{\eta\rightarrow 0}\Im\left[\omega - {\cal H}_{k_x}+i\eta\right]^{-1}_{yy},
\end{align}
defined for a nanoribbon structure \cite{LiuPRB2021}. It is directly related to the dynamical structure factor measured in neutron-scattering experiments \cite{mourigal2010field,zhitomirsky2013colloquium}. 
\begin{figure}
    \centering
    \includegraphics[width = 8.5cm]{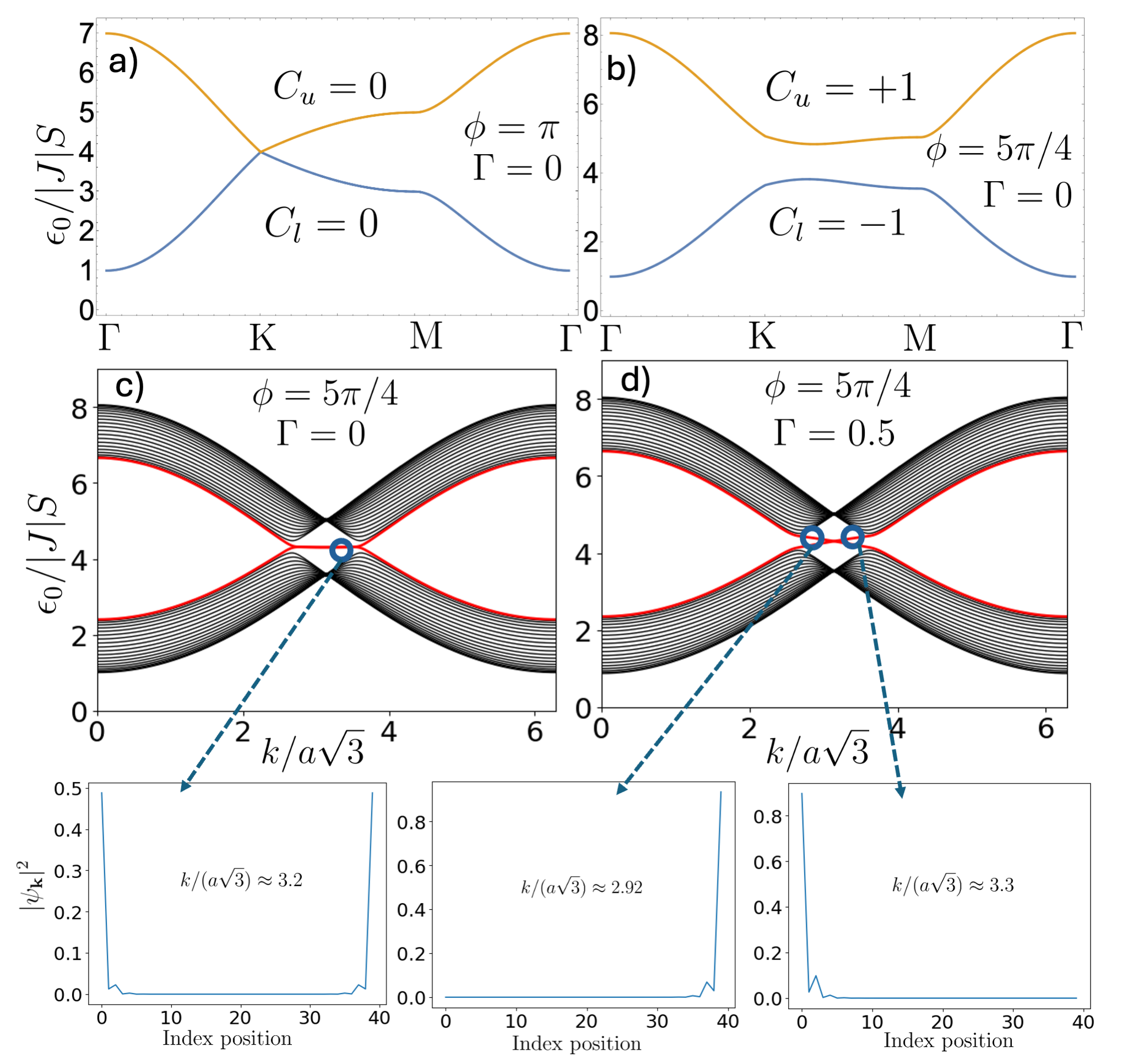}
    \caption{Undeformed magnon energy and spectral function in a H-K magnet for $\Gamma = 0.0$ with a) $\phi = \pi$ and b) $\phi=5\pi/4$. c) and d) depict the magnon bands along the zigzag edge in a nanoribbon structure with $\phi = 5\pi/4$ for the case c) $\Gamma =0$ and d) $\Gamma = 0.5$, respectively. In the insets, we show the square module of the magnon wave function as a function of the position index $y$ for the depicted wave vector.  e) and f) correspond to the bulk spectral function of c) and d), respectively. All calculations consider an out-of-plane magnetic field $B=1$, $S=J=1$, and the nanoribbon is composed of $N_y=20$ unit cells.}
    \label{fig:fig2}
\end{figure}

Magnonic fluctuations of the ferromagnetic state in the unstrained HK model exhibits topological phases \cite{joshi2018topological,aguilera2020topological}. The spectrum of the two-band magnon Hamiltonian, $\epsilon_{\pm}({\bs k})$ is shown at Fig. \ref{fig:fig2} along high-symmetry points. Setting $\Gamma = 0$, we consider $\phi = \pi$ and $\phi=5\pi/4$ at panels (a) and (b), respectively. The topological gap at the Dirac point K (K') is determined by the Kitaev term $K$ and the Chern number for the upper and lower band is $C_{u}=-C_{l}=1$ \cite{joshi2018topological,aguilera2020topological} (see Fig. \ref{fig:fig2}b)). For a strip geometry, with zig-zag edges and finite size along $y$-axis, and using periodic boundary conditions along $x$ direction, we plot the spectrum at Figs. \ref{fig:fig2}c) and \ref{fig:fig2}d), for $\Gamma=0$ and $\Gamma=0.5$, respectively. Note that the edge modes highlighted in red are non-dispersive near the Dirac point for $\Gamma=0$. The topologically protected states are localized at the boundaries, as demonstrated in the insets of Figs. \ref{fig:fig2}c) and \ref{fig:fig2}d), where we depict the square module of the magnon wave-function $\psi_{\bm k}$ as a function of the position index $y$.

\begin{figure*}
    \centering
    \includegraphics[width = 17.5cm]{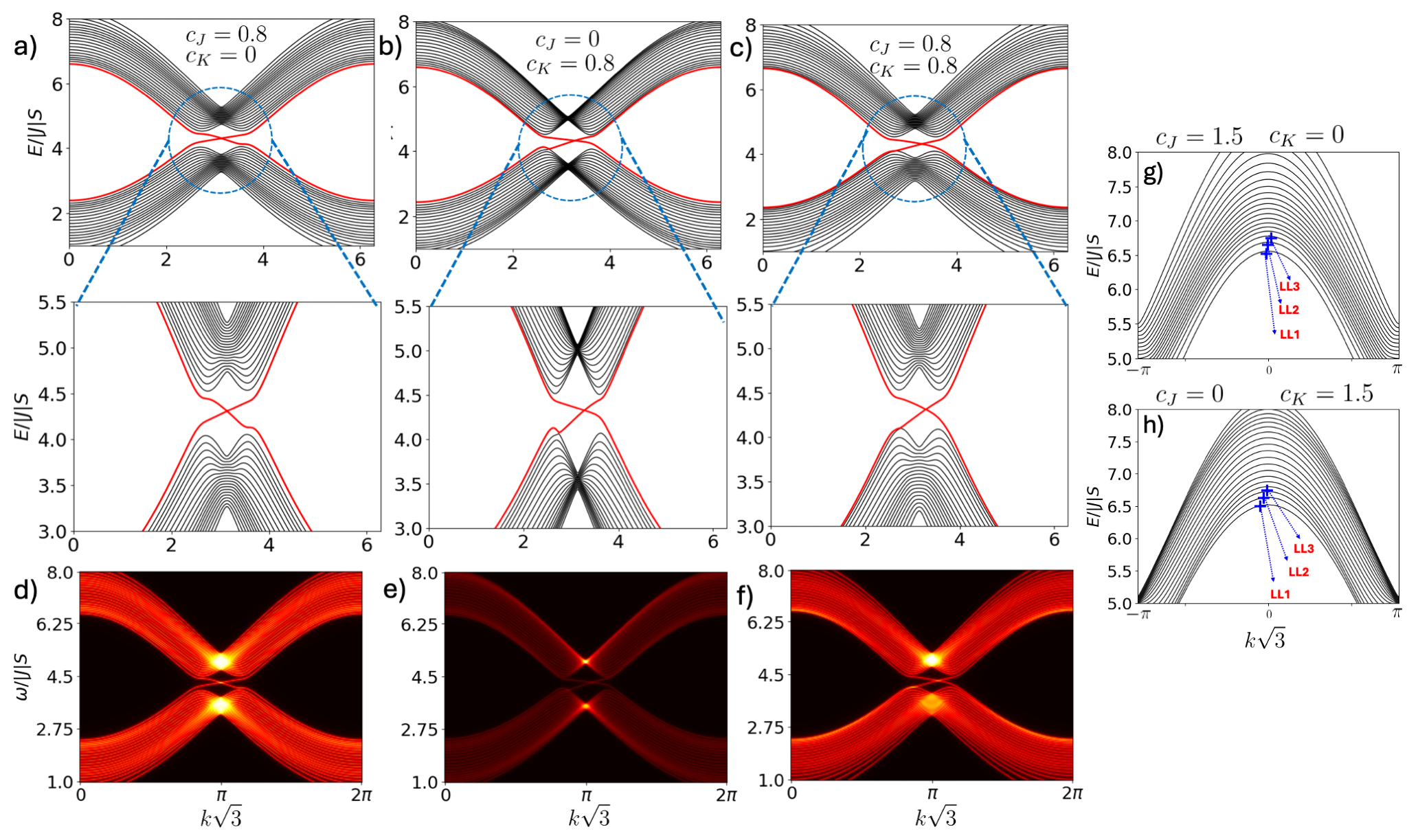}
    \caption{Deformed H-K nanoribbon under the application of an uniaxial strain for $\phi=5\pi/4$, $\Gamma = 0.5$ and an external magnetic field $B=1$. a) depicts the case with $c_J=0.8$ and $c_K=0$; b) $c_J=0$ and $c_K=0.8$; and c) $c_J=0.8$ and $c_K=0.8$. Panels in the middle show a zoom of the in-gap edge states for panels a), b), and c), respectively. Their corresponding spectral function is shown in panels d), e), and f). The column at the right (panels g) and h)) shows a zoom-in to the upper bulk bands around $k=0$ of the deformed H-K magnet ($c_J=1.5$ with $c_K=0$ and $c_J=0$ with $c_K=1.5$) depicting the generated magnon Landau Levels (LL). 
    In all cases, it was considered $N_y= 20$.}
    \label{fig:fig3}
\end{figure*}

\section{Strained lattices}
We now study how the band structure is affected by the presence of distinct strains. We will consider a semi-infinite system, with periodic boundary conditions along the $x-$axis and opened along the $y-$axis. In our numerical calculations, we will adopt $\hbar=1$, the spin length $S=1$, the lattice constant $a_0=1$, and an external magnetic field $B=1$. In addition, we will consider the case $\Gamma = 0.5$ and $\phi = 4\pi/5$ as we are primarily interested in the behavior of topological edge modes under the action of external strains. Nevertheless, by considering different values for $\Gamma$ and $\phi$, it gives rise to distinct magnetic ground states. In such a case, our model still holds but a local spin rotation on every site must be performed to explore magnons lying in the corresponding magnetic texture. 
\subsection{Uniaxial deformation}
We consider the crystal lattice under uniaxial strain along the $y-$direction, and restrict to deformations quadratically dependent on position $y$, which means that the strain tensor varies linearly on $y$. Thus, the magnetic couplings reads $\mathcal{X}^{l}_{ij} = \mathcal{X}-\gamma^{ij}_{\eta} u_{ij}=\mathcal{X}-\gamma^{ij}_{\eta}c_{\chi} y$, where $c_{\chi}\gamma^{ij}_{\eta}=c_{\chi}\delta_{\eta}^i\delta_{\eta}^j$ comprises the magnetoelastic coupling relative to the three distinct links labeled by $\eta$, and $c_{\chi}$ controls the strain strength. Setting the origin at the bottom of the ribbon, the vertical maximum position is given by $y_{\text{max}} = {3}(N_y -1)/{2}+{1}/{2}$, being $N_y$ the number of sites along the $y-$ direction. We consider that spins located at the upper edge of the ribbon do not interact when subjected to a large strain. This fact defines the maximum strength strain by demanding that $\mathcal{X}-\gamma^{ij}_{\eta} u_{ij} = \mathcal{X}-c_{\text{max}}y_{\text{max}}=0$. Thus, we set $c_{\text{max}}=1/y_{\text{max}}$, which could be considered as the scale for the control parameter $c_{\chi}$. Including $\mathcal{X}^{l}_{ij} $ into the magnon Hamiltonian, we determine the energy spectrum numerically by employing the Colpa algorithm \cite{colpa1978diagonalization}. 

In Fig. \ref{fig:fig3}, we show the results where $c_J$ and $c_K$ encode the Heisenberg and Kitaev coupling modulations, respectively. As we show below, unaxial deformations affect mainly edge magnons located around the boundary of the BZ ($k\sqrt{3}\approx \pi$), so we show the magnons bands in the range $[0,2\pi]$. In panels (a), (b), and (c), the energy spectrum is displayed for $c_J=0.8$ with $c_K=0$;  $c_J=0$ with $c_K=0.8$; and $c_J=c_K=0.8$, respectively. From Fig. \ref{fig:fig3}a), it can be seen that the modulation over the $J$ parameter slightly distorts the bulk band structure, and the major changes are mainly appreciable in the in-gap magnon states around $k\sqrt{3}=\pi$ (recall $a_0 = 1$), as depicted in the second row. Indeed, one can see that the crossing point between the two edge modes is shifted to lower momenta and the shape of both edge modes is differently affected (see Fig. \ref{fig:fig4} below). Fig. \ref{fig:fig3}d) depicts the magnon spectral function, where the brighter zones mean larger available magnon states. In this case, the upper and lower bulk bands show a similar behavior around $k\sqrt{3}=\pi$, so the sole application of $c_J$ has consequences mainly on the shape of the edge states instead affecting the general form of bulk bands. When turning off the modulation of the Heisenberg parameter and only allowing the strain-mediated modulation of the Kitaev term, the shift of the crossing points is inverted toward higher momenta, as depicted in Fig. \ref{fig:fig3}b) and the corresponding zoom at the middle row. In this case, the crossing point of the edge bands is shifted toward higher momenta, and the spectral function shows a pronounced decay in its amplitude, meaning a diminishing of the available magnon states. Notably, one of the in-gap edge states is deformed and tends to hybridize with the topologically trivial bulk bands. We consider this behavior as a manifestation of the topological nature of magnon edge bands in the presence of the Kitaev terms. While the deformation fields promote an hybridization of such edge states with the bulk ones, the topological protection still preserves, avoiding such an hybridization through the formation of new band gap. This effect is more appreciable for larger strengths strains $c_K$ (see Fig. \ref{fig:fig4}).

Next, in Fig. \ref{fig:fig3}c), we show the case where both Heisenberg and Kitaev parameters are modulated by strain. As can be seen, the effect is now moderated because it results from the competition between the both phenomena presented above. Nevertheless, due to the fact we are dealing with linear strain fields, we expect the inverse behavior when changing the sign of $c_K$ and $c_J$. Therefore, it should be possible to enhance or moderate the tendency of the edge-modes to hybridize with the bulk ones. Finally, the magnon spectral function has a notorious amplitude increase in the upper bulk bands, which we attribute to the combination of the pseudo-magnetic field induced by nonuniform deformations together with the external magnetic field. This fact is supported by the presence of magnon Landau Levels (LL), which we show in Figs. \ref{fig:fig3}g) and \ref{fig:fig3}h) for selected values of $c_J$ and $c_K$. The pseudo magnetic field emerges from the gauge field induced by nonuniform deformation fields, as in the case presented here. Thus, the analysis of the spectral function allows us to conclude that considering $c_J$ or $c_K$ separately produces slight changes in the magnon population at the boundary of the BZ. However, when combined, one can appreciate a marked difference in the spectral function between the bulk bands of higher and lower energies, where a larger magnon population for higher energy bands is evidenced. The modification of the magnon spectral function mediated by external strains might have several consequences in practical applications related to the magnon density of states and thermodynamic processes \cite{vidal2024magnonic}.

The application of uniaxial strains also impacts the magnon velocity. In Fig. \ref{fig:fig4}, we show the magnon velocity $v_m$ of both magnon edge states for different strain strengths, $c_J$, and $c_K$, as a function of the wave vector, in the vicinity of $k\sqrt{3}=\pi$. The solid line corresponds to the in-gap mode that we called \textit{upper state} (since it localizes at the lower end in the nanoribbon), while dashed lines are the ones representing the \textit{lower states} (see the band structure in the zoom-in at the insets). Fig. \ref{fig:fig4}a) shows the case for $c_K=0$ and distinct values of $c_J$, so it focuses on the sole effect of deformations affecting the exchange interaction. As can be seen, the difference in the magnon velocity between both in-gap modes depends on the strain strength $c_J$. Notoriously, the magnon velocity of upper states is more affected by the uniaxial strain, and such an effect is more evident for larger strains. This is consistent with the shape that topological magnon bands acquire in Fig. \ref{fig:fig3}a), where a greater slope of the corresponding magnon band is observed for a given strain strength. Next, we focus on the sole effect of considering $c_K\neq 0$, which is shown in Fig. \ref{fig:fig4}b). It can be noticed that there is a marked difference between the magnon velocity for the upper and lower states. Indeed, for the lower states, there is a notorious change in the magnon velocity, which comes from the band deformation depicted in Fig. \ref{fig:fig3}b). Such a behavior becomes more intense as the strain strength is larger. As stated above, this results from avoiding the hybridization between the in-gap state with the bulk ones, which we interpret as a manifestation of the topological protection. Finally, we combine both cases presented above and, in Fig. \ref{fig:fig4}c), we show results of the magnon velocity for $c_K = c_J\neq 0$. One can observe that the separated effects of $c_J$ and $c_K$ are now in competition, so the changes in magnon velocity are moderated. Nevertheless, we still notice that the magnon velocity for both in-gap states is essentially different according to the applied strain. Thus, we conclude that magnons become strongly non-reciprocal around $k\sqrt{3}=\pi$, and such a nonreciprocity depends on the strain strength. Interestingly, the magnon nonreciprocity at the boundary of BZ might produce another interesting effect, such as changes in the magnon localization. Indeed, as long the in-gap magnon bands cross at a different k-point compared with the undeformed case, the topological nature of such magnon bands allows a distinct localization of magnon states mediated by strain. In the case of the presented uniaxial strain, such an effect is slight and not shown here. However, as we see in the next section, applying a different strain could make the effect more appreciable.
\begin{figure}
   \includegraphics[width=8.3cm]{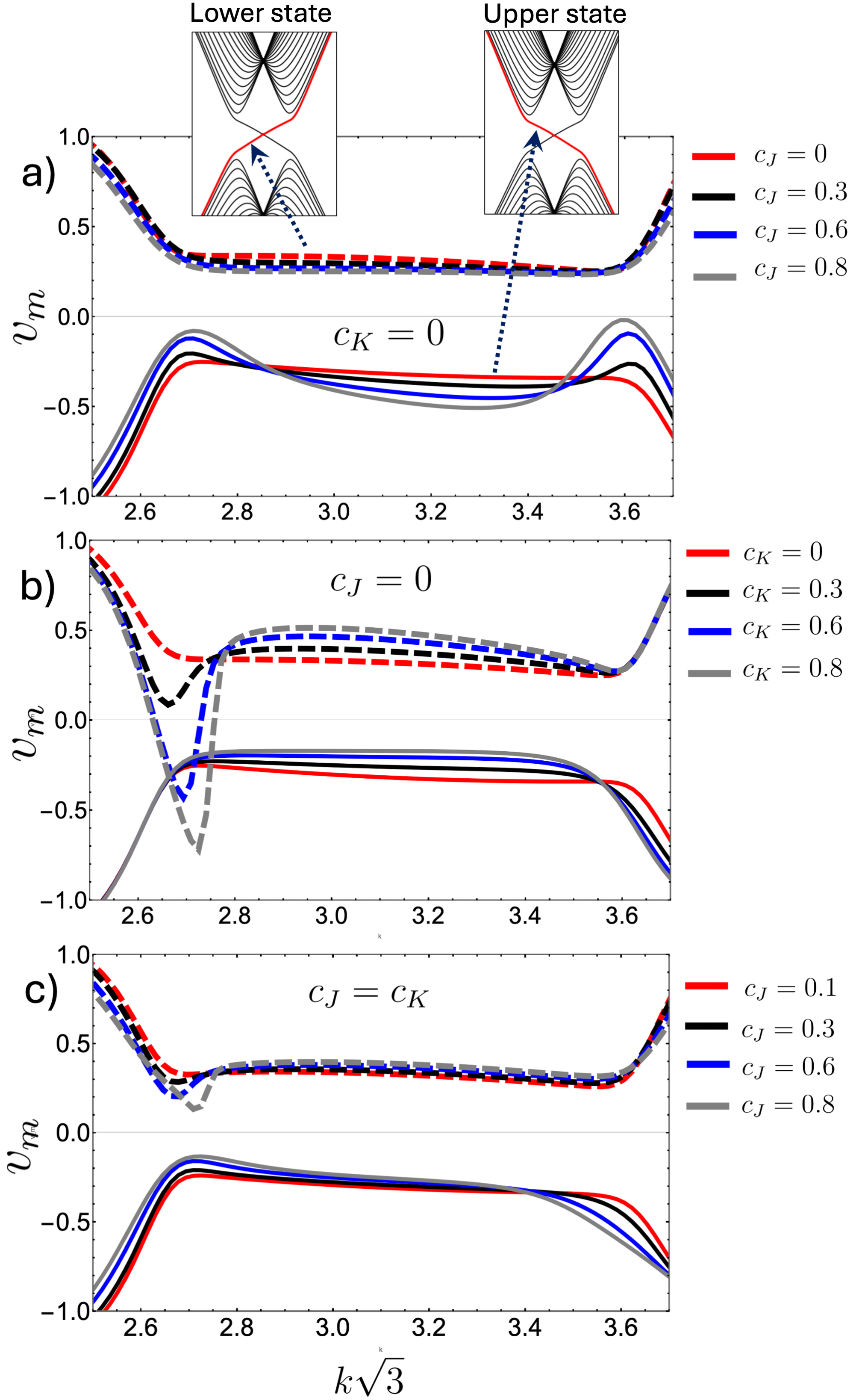}
    \caption{Magnon group velocity around $k\sqrt{3}=\pi$ for the topologically protected edge-states with different strain strengths. Panel (a) corresponds to the case $c_K=0$ and different values of $c_J$. Panel (b) corresponds to $c_J=0$ and different values of $c_K$, while panel (c) shows the case $c_J=c_K$. The dashed and solid lines illustrate the corresponding group velocity of the topological bands depicted in panel (a), respectively. The names "upper" and "lower" states refer to the localization of magnon states in the ribbon.}
    \label{fig:fig4}
\end{figure}

\subsection{Twisted lattice deformation}
We now explore the effects on the magnon band structure when the lattice system is twisted by the deformation field ${\bs u}(\bm{r})=(0,y\cos\lambda x-y,y\sin\lambda x)$. For a nanoribbon with dimensions $L\times W$, the twist is quantified by an angle $\Omega$, see Fig. \ref{fig:fig1}(c), so that $\lambda=\Omega/L$ and where it is assumed $L\gg W$. Under strain the length of bonds become $\tilde{\delta}_i = \sqrt{\delta_i^2 + \lambda^2\delta_{i,x}^2(y^2+y\delta_{i,y})}$, which in turn modulates the nearest neighbor interactions. The magnetic couplings, assumed to be exponentially decaying $\mathcal{X}_i = \mathcal{X} \exp{(-\beta(\tilde{\delta}_i-\delta_i)/\delta_i)}$ with $\beta\approx 1$ the Gruneisen parameter, becomes
\begin{align}
    \mathcal{X}_i=\mathcal{X}_i\exp{\left[1-\sqrt{1+\frac{3}{4}\lambda^2(y^2+\frac{a}{2}y)}\right]}.
    \label{eq:expcoupling}
\end{align}
Assuming a small twist we expand Eq. (\ref{eq:expcoupling}), finding that only bonds $x$ and $y$ are affected, so that
\begin{align} \mathcal{X}_x=\mathcal{X}_y=\mathcal{X}-c \left(y^2+\frac{a}{2}y\right), \hspace{0.3cm}  \mathcal{X}_z=\mathcal{X},
\label{eq:expcoupling2}
\end{align}
where $c = \frac{3}{8}\lambda^2\mathcal{X} $. We introduce Eq. \eqref{eq:expcoupling2} into the magnon Hamiltonian and numerically diagonalize it for different deformation strengths $\lambda$. In addition, we consider that Heisenberg and Kitaev couplings are separately deformed by a parameter $\lambda_{J,K}$. 

\begin{figure*}
   \includegraphics[width=15cm]{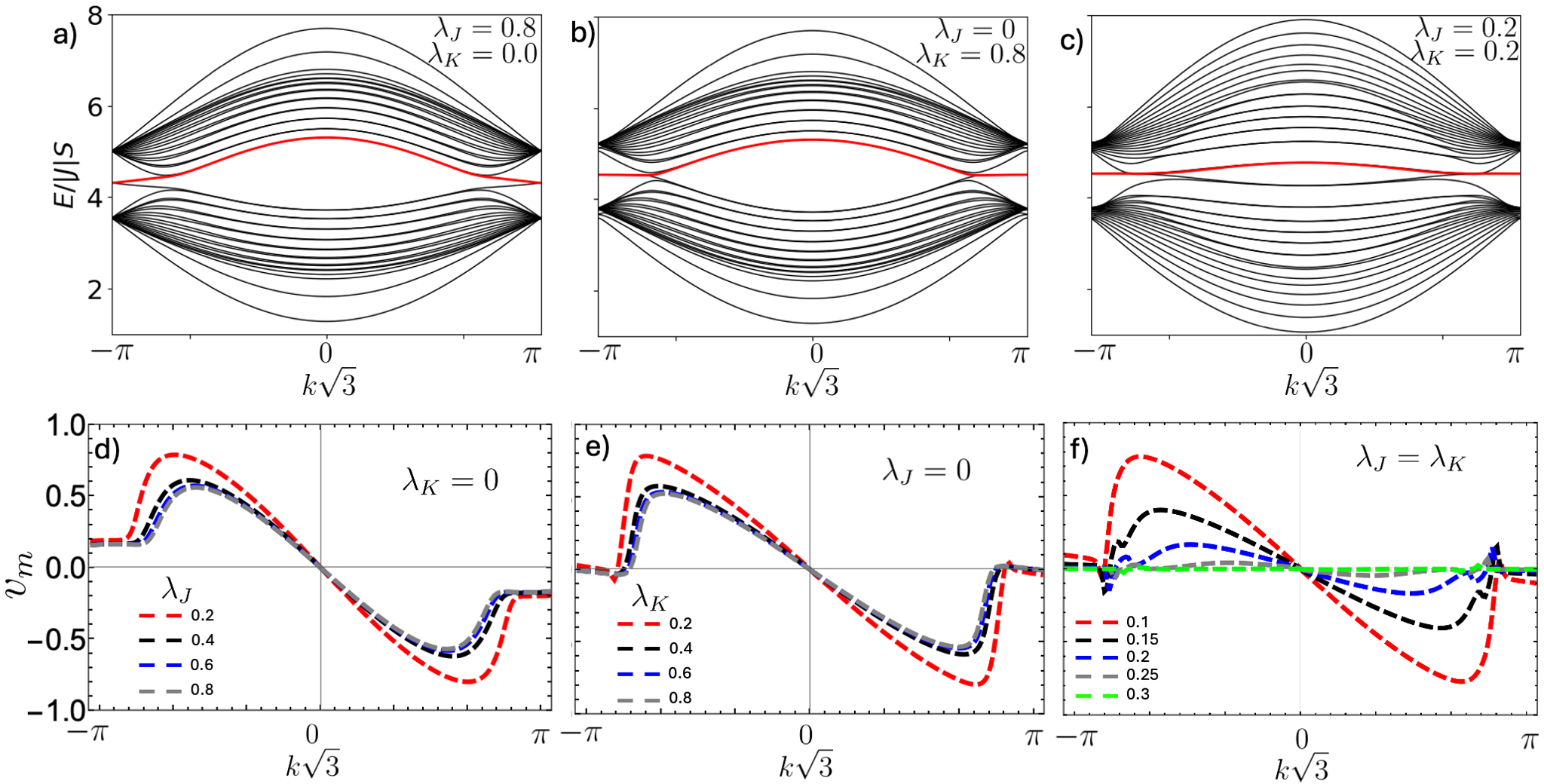}
    \caption{(Upper) Twisted H-K nanoribbon for $\phi=5\pi/4$, $\Gamma = 0.5$ and an external magnetic field $B=1$. a) depicts the case with $\lambda_J=0.8$ and $\lambda_K=0.0$; b) depicts the case $\lambda_J=0$ and $\lambda_K=085$; and c) corresponds to the case $\lambda_J=\lambda_K=0.2$. (Lower) Magnon group velocity for the corresponding magnon band marked in red at the upper panel for different values of $\lambda_J$ and $\lambda_K$. In d) we show the magnon group velocity for the case $\lambda_K=0$ and different values of $\lambda_J$. e) depicts the case $\lambda_J=0$ and different values of $\lambda_K$, and f) corresponds to the case with $\lambda_J=\lambda_K\neq 0$}
    \label{fig:fig5}
\end{figure*}

Fig. \ref{fig:fig5} shows the twisted nanoribbon for different values of $\lambda_{J}$ and $\lambda_{K}$. Unlike the uniaxial deformation, the twist one significantly impacts the entire Brillouin zone, so we depict the magnon bands in the range $[-\pi,\pi]$. 
Figs. \ref{fig:fig5}a) -\ref{fig:fig5}c) show the twisted nanoribbon $\lambda_J=0.8$ and $\lambda_K=0$,$\lambda_J=0$ and $\lambda_K=0.8$, and $\lambda_J=\lambda_K=0.2$, respectively. One can appreciate that the sole presence of $\lambda_J$ (Fig. \ref{fig:fig5}a)) mainly contributes to making the bands more dispersive, in agreement with the expected presence of magnon Landau Levels for nonuniform strain fields, while the general shape of one of the edge modes (marked in red) remains almost unaltered. On the other hand, when considering only $\lambda_K$, the presence of magnon LL is also observed, but, interestingly, the edge modes are flattened at the boundary of BZ, as shown in Fig. \ref{fig:fig5}b). A notable behavior is observed when both $\lambda_K$ and $\lambda_K$ are no null. Indeed, Fig. \ref{fig:fig5}c) shows that the combination of the two effects mentioned above gives rise to a flattening of the magnon edge bands. This effect is better appreciable in Figs. \ref{fig:fig5}d) - \ref{fig:fig5}f), where we show the magnon group velocity $v_m$ for the cases presented at the upper panel but varying the strength of the deformation parameter $\lambda_{J,K}$. As can be seen, while the effect of $\lambda_J$ is related to a constant magnon velocity at the boundary of BZ (\ref{fig:fig5}d)), the effect of $\lambda_K$ is clearly making magnons non-propagative since $v_m\approx 0$ at such wavevectors, as shown in Fig. \ref{fig:fig5}e). Note that away from the boundary of BZ, the magnon velocity does not present major differences when considering the separated effects of $\lambda_K$ and $\lambda_J$. However, as observed in Fig. \ref{fig:fig5}f), the combined effect of both deformation parameters allows the formation of flat bands along the entire Brillouin zone. This is essentially different from previous studies, where the main focus was on the behavior of magnons around the boundary of BZ or around Dirac points. Here, we show the possibility of having non-propagative magnons along the entire BZ, which indeed depends on the deformation strength parameters.

\begin{figure}
   \includegraphics[width=8.3cm]{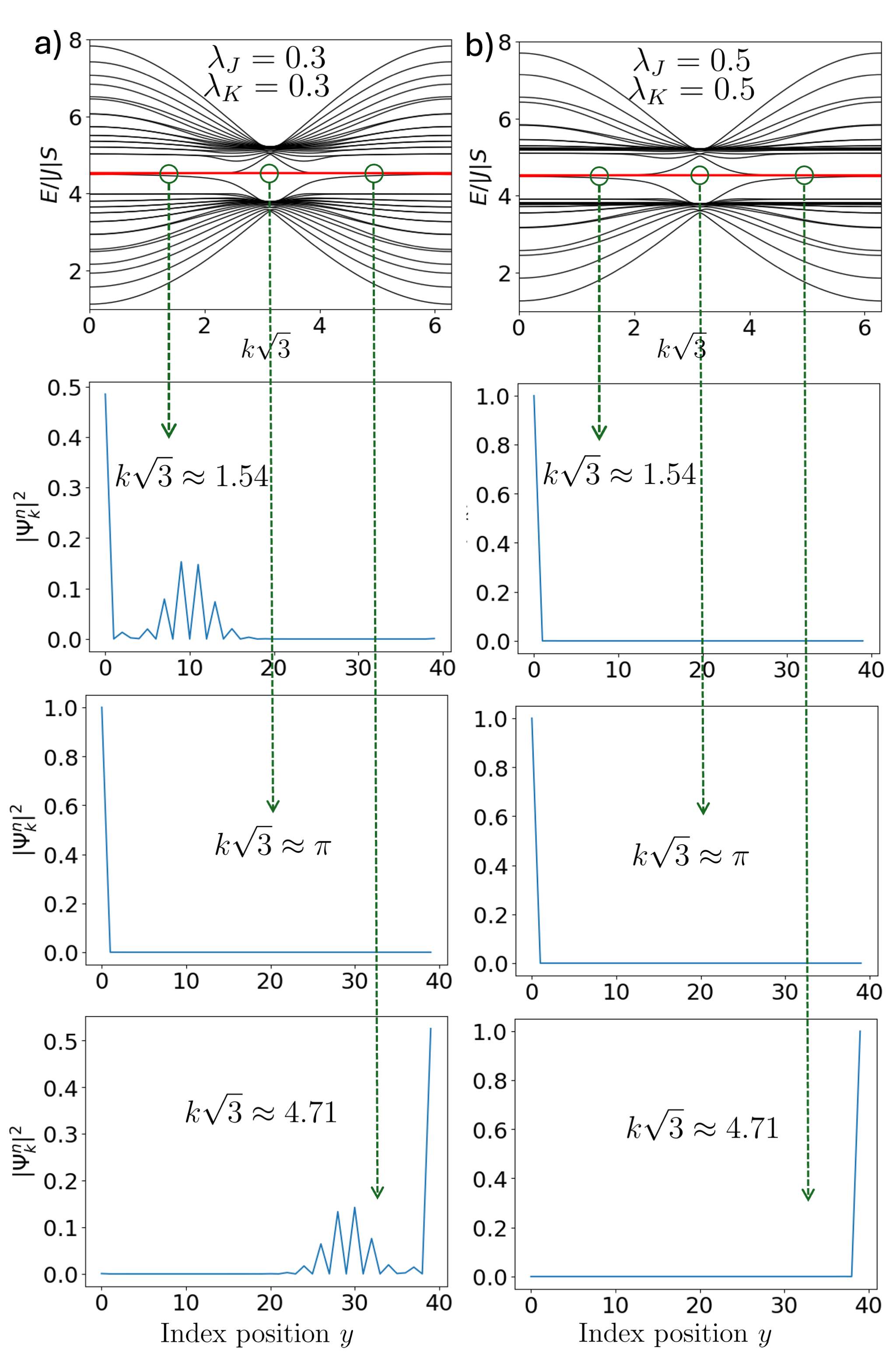}
    \caption{(Upper) Magnon bands of the twisted H-K nanoribbon for $\phi=5\pi/4$, $\Gamma = 0.5$ and an external magnetic field $B=1$. a) depicts the case with $\lambda_J=\lambda_K=0.3$ and b) $\lambda_J=\lambda_J=0.5$. The three panels below a) and b) correspond to the square module of the magnon wavefunction for an in-gap magnon mode as a function of the index position $y$ at three different wave vectors. The arrows show at which wavevector the magnon localization is computed. }
    \label{fig:fig6}
\end{figure}

\begin{figure}
   \includegraphics[width=8.3cm]{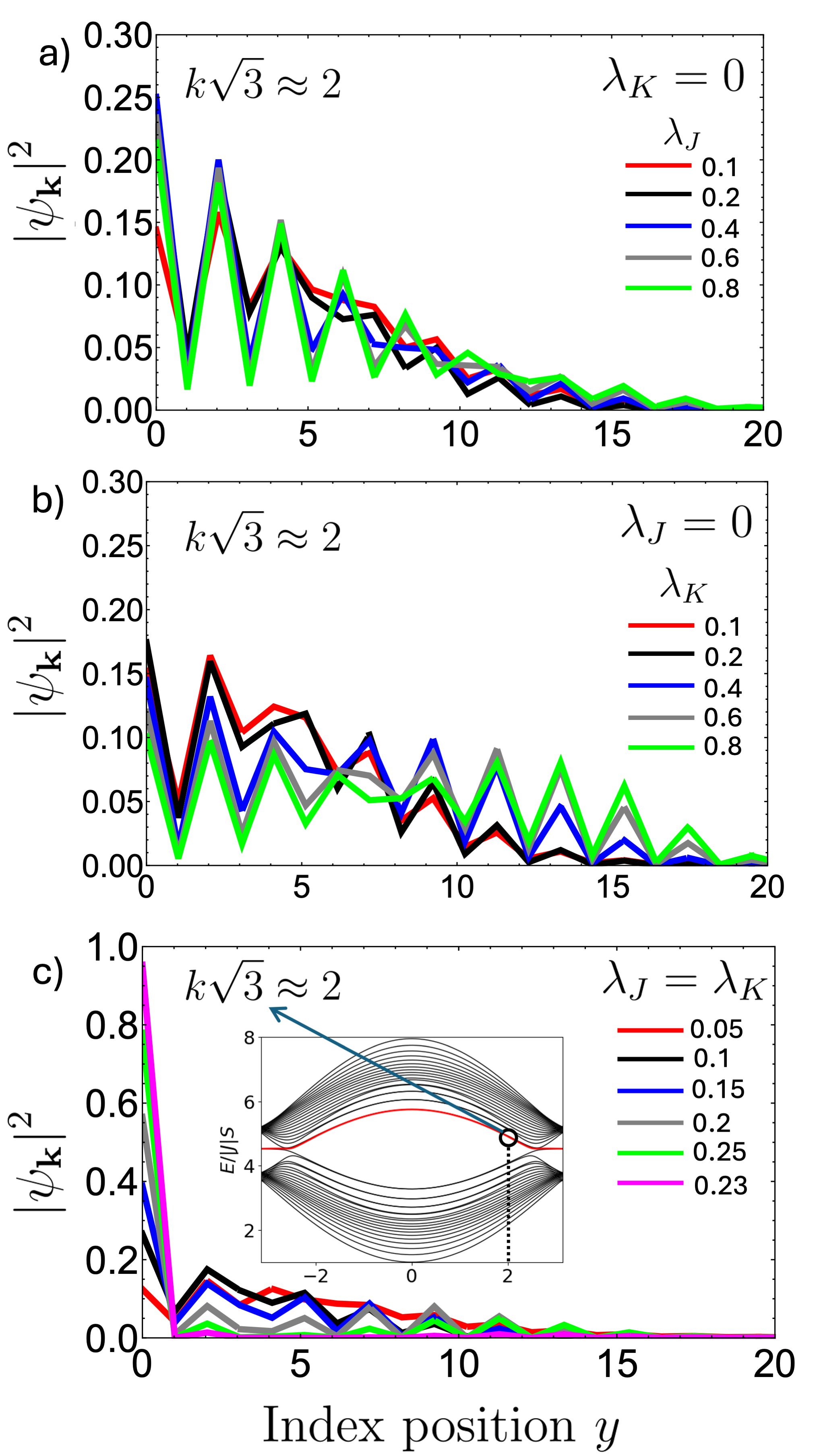}
    \caption{Probability of magnon localization $\vert\psi_{\mathbf{k}}\vert^2$ as a function of the index position $y$ with a fixed band and wave vector $k\sqrt{3}\approx 2$ for different values of the twist parameters $\lambda_{J,K}$. a) depicts the case $\lambda_K=0$, b) the case $\lambda_J=0$ and c) corresponds to $\lambda_J=\lambda_K$. The inset of magnon bands at panel c) is shown as a reference of the specific wave vector value used and the corresponding band highlighted in red.}
    \label{fig:fig7}
\end{figure}

Following previous analysis, in Fig. \ref{fig:fig6} we show the behavior of magnon edge modes for selected values of $\lambda_K=\lambda_J\neq 0$. As stated above, this combination gives rise to flat bands that encode non-propagative magnons. Since magnon in-gap modes have a topological origin, the non-propagative (topological) magnon modes induced by the twist deformation might still preserve their topological features, such as the localization at the boundaries of the sample. In the left panel (Fig. \ref{fig:fig6}a)), we show the case for $\lambda_J=\lambda_K=0.3$, while Fig. \ref{fig:fig6}b) corresponds to the case $\lambda_J=\lambda_K=0.5$. The lower panels in both cases correspond to the probability of finding an (edge) magnon state as a function of the index position $y$, $\vert\psi_{\mathbf{k}}\vert^2$. The position $y=0$ ($y=40$) corresponds to the lower (upper) nanoribbon end. We show such a probability for three different values of the wave vector ($k\sqrt{3}\approx 1.54, \pi$ and $4.71$). As can be seen, for the case $\lambda_J=\lambda_K=0.3$, \textit{edge} magnons are hybridized with bulk ones at $k\sqrt{3}\approx 1.54$ since the probability of finding magnons at the lower ends and at the bulk of nanoribbon is nonzero. A similar behavior occurs for $k\sqrt{3}\approx 4.71$, but bulk magnons are now mainly hybridized with \textit{edge} magnons of the upper ends. However, for $k\sqrt{3}\approx \pi$, magnons are completely localized at the lower end of the nanoribbon as they are essentially topologically protected magnons. However, such localization can indeed be controlled by the strength of the deformation parameter $\lambda_{J,K}$. Fig.  \ref{fig:fig6}b) shows the probability of finding a magnon state as a function of the index position $y$ for the same wave vectors presented in Fig. \ref{fig:fig6}a) but with a larger deformation parameter, i.e., $\lambda_J=\lambda_K=0.5$. In this case, magnons are always located at one of the ends of the nanoribbon, and no hybridization with magnons lying at bulk bands is observed. This feature invites the possibility of controlling the localization of the magnon edge states with the deformation parameter $\lambda_{J,K}$. In this regard, in Fig. \ref{fig:fig7}, we show the probability of finding an (edge) magnon state at $k\sqrt{3}\approx 2$ as a function of the index position $y$ for the magnon band marked in red in the inset of Fig. \ref{fig:fig7}c) and different values of $\lambda_{J,K}$. Panels a), b), and c) show the cases with $\lambda_K=0$, $\lambda_J=0$, and $\lambda_K=\lambda_J$, respectively.  As can be seen from panels a) and b), the separated effect of $\lambda_J$ and $\lambda_K$ on magnon bands does not allow the controlled localization of magnon at the (lower) end of the nanoribbon as they do not generate flat bands. Indeed, these cases show edge magnons mainly hybridized with bulk ones. However, as depicted in Fig. \ref{fig:fig7}c), as long $\lambda_{J,K}$ becomes larger, it is possible to localize the corresponding magnons at the lower end of the nanoribbon. As stated above, this behavior originated from the formation of flat bands due to the application of strains that modulate both $J$ and $K$ through $\lambda_{J,K}$.

\section{Conclusions and discussions}
We have explored the magnon properties in strained H-K magnets. Since magnon excitations in these materials are genuinely topological, we focused on the edge modes and evaluated the effect of strain on them. By expanding the relevant magnetic parameters as a function of the position, we have presented a theory that accounts for weak external strains, which is valid along the whole Brillouin zone. Our results showed that, in general, weakly strained H-K magnets are modeled by a Hamiltonian non-local in the momentum space. Such a constraint can be overcome by considering linear deformation fields, making the magnon Hamiltonian become local in the momentum space so that, in principle, it can be exactly diagonalizable for an infinite sample.

We test our results in semi-infinite nanoribbons by applying two kinds of strains, namely, uniaxial and twist deformations. For the former, we found that magnon bands become more dispersive, in concordance with previous results related to the presence of magnon Landau levels. Also, the magnon group velocity of edge modes could suffer from extensive changes according to the strength strain. Indeed, for a given combination of Kitaev- and exchange-mediated strain parameters, we found a non-reciprocal behavior between the in-gap edge modes, which we attributed to the linear dependence of the strain field into the Hamiltonian. Such a non-reciprocity depends on the strain strength $c_{J,K}$. Based on it, we predict a control over the magnon group velocity according to the sign of the strain parameter (which codes a stretch or elongation strain). Thus, we have shown that linear deformation fields can give rise to a non-reciprocal behavior for in-gap edge magnon modes, which allows for controlling the magnon velocity and the corresponding non-reciprocity.

For twist deformations, we found that bulk magnon bands also become more dispersive, which is a consequence of the presence of predicted magnon Landu Levels, and the separated effect of Kitaev- and exchange-mediated strain parameters promotes different results. Specifically, the sole strained exchange interaction mediates mainly the band dispersion, while the strained Kitaev interaction is responsible for the flattening of the magnon edge modes around the boundary of the Brillouin zone. However, when both act simultaneously, the combined effect gives rise to flat bands with non-propagative in-gap and bulk modes, which eventually hold along the whole Brillouin zone according to the strain strength. Notoriously, the topological protection of in-gap modes holds even far from the boundary of BZ, as demonstrated by the localization of magnon modes at different wavevectors. Thus, we proved the control of non-propagative and robust magnonic edge modes.

Finally, both uniaxial and twist deformations have major consequences on the magnonic edge modes. Since such modes are topologically non-trivial and whose presence depends on the Kitaev interaction, we claim that both the non-reciprocal behavior (for uniaxial strain) and the presence of flat bands (for twist deformations), which are controlled by the strain strength, might have a deeper incidence on the topological protection of the in-gap modes. Indeed, although we have demonstrated that the characteristic localization of topological magnons holds for the distinct strains considered here, a comprehensive study on topological invariants and/or the effect of disorder on the magnon edge modes in the presence of different strains is needed to get further conclusions. Thus, such a task is left for future work.

The main contribution of our work relies on the fact that several properties of magnons hosted in strained H-K can be controlled according to the type of the applied strain. Unlike previous works on strained magnets, our work goes beyond the expansion around Dirac points and presents a theory fully applicable along the whole Brillouin zone. Although we have performed the calculations in a 2D-ferromagnetic material, our model is general and supports different magnetic phases since the inclusion of strains does not depend on the specific magnetic configuration. In addition, we considered magnetic parameters relative to a magnetic monolayer accounting for only exchange and Kitaev interaction. However, symmetry arguments also allow Dzyaloshinskii-Moriya (DM) interaction. We have disregarded its contribution because we were mainly interested in the effect of modulating the Kitaev parameter. Nevertheless, the same methodology can be employed to explore possible effects arising from the DM interaction in the presence of strain.

\section{Acknowledgments}
R.E.T thanks funding from Fondecyt Regular 1230747 and N.V-S. thanks funding from Fondecyt Regular 1250364 

\bibliography{strainKitaev}
\section{Appendix}

The magnon spectrum for $K^{\gamma}=K$ in the presence of an external magnetic field $B$ reads
\begin{align}
\epsilon^2_{\pm}(\bm k) = f_0+f_1(\bm k)\pm\sqrt{f_2(\bm k)},
\end{align}
where $f_0=S^2[4(J+K)(3J+K) -(\Gamma^z)^2+B^2-2B(3J+2K)]$, while  $f_1(\bm{k})$ and $f_2(\bm{k})$ reads: 


\begin{multline}
\frac{f_1(\bm{k})}{2S^2(J+K)} = (J+K) \cos \left(\frac{1}{2} (\sqrt{3} k_x - 3 k_y)\right) + \\
J \left[\cos \left(\frac{1}{2} (\sqrt{3} k_x + 3 k_y)\right) + \cos \left(\sqrt{3} k_x\right)\right]
\end{multline}

\begin{widetext}
\begin{align}
\frac{f_2(\bm{k})}{2S^4} = &\;\nonumber
2 (B - 3J - K)(3J^2 + 4JK + 2K^2)(B - 3(J+K)) 
- 2J(J+2K)\Gamma_z^2 \\\nonumber
&+ 4J(J+K)(-B+3J+K)(-B+3(J+K)) \cos\left( \sqrt{3}k_x \right) \\ \nonumber
&+ J\left( JK^2 + (J+2K)\Gamma_z^2 \right) \cos\left( 2\sqrt{3}k_x \right) \\\nonumber
&+ 4 \cos\left( \frac{\sqrt{3}}{2}k_x \right)
\Big[
18J^4 + 51J^3K + 52J^2K^2 + 22JK^3 + 4K^4 \\\nonumber
&\quad + B^2(J+K)(2J+K)
- 2B(J+K)(2J+K)(3J+2K) \\\nonumber
&\quad - (J^2+2JK+2K^2)\Gamma_z^2
+ \left( JK^2(J+2K) + (J^2+2JK+2K^2)\Gamma_z^2 \right) \cos\left( \sqrt{3}k_x \right)
\Big] \cos\left( \frac{3}{2}k_y \right) \\\nonumber
&+ \left[
4JK^2(J+K)
+ \left( K^2(5J^2+8JK+4K^2) + J(J+2K)\Gamma_z^2 \right) \cos\left( \sqrt{3}k_x \right)
\right] \cos\left( 3k_y \right) \\\nonumber
&- 4 \sin\left( \frac{\sqrt{3}}{2}k_x \right) \sin\left( \frac{3}{2}k_y \right)
\Big[
K\left( (B-3J)^2J + (B-6J)(B-4J)K + 2(-2B+9J)K^2 + 4K^3 \right) \\\nonumber
&\quad - (J^2+2JK+2K^2)\Gamma_z^2
+ \left( JK^2(3J+2K) - (J^2+2JK+2K^2)\Gamma_z^2 \right) \cos\left( \sqrt{3}k_x \right)
\Big] \\
&- (J+2K)(3JK^2+2K^3-J\Gamma_z^2) \sin\left( \sqrt{3}k_x \right) \sin\left( 3k_y \right)
\end{align}
\end{widetext}
\end{document}